# Trainmon: a framework for reverse engineering potentials in superconducting Qubits

Saeed Hajihosseini[1], Seyed Iman Mirzaei[2], Hesam Zandi[1,3,4], Mohsen Akbari[5]

[1] Iranian Quantum Technologies Research Center (IQTEC), Tehran, Iran

[2] Department of Condensed Matter Physic, Faculty of Basic Sciences, Tarbiat Modares University, Tehran, Iran

[3] Faculty of Electrical Engineering, K. N. Toosi University of Technology, Tehran, Iran

[4] Electronic Materials Laboratory, K. N. Toosi University of Technology, Tehran, Iran

[5] Quantum optics lab, Department of Physics, Kharazmi University, Tehran, Iran

## Abstract

A framework named Trainmon is introduced to reverse-engineer a quantum potential well for superconducting qubits. Trainmon consists of parallel branches of identical Josephson junctions. The Hamiltonian for this circuit resembles a discrete cosine transform, which can be applied to mimic various potentials. This framework is applied to well-known qubit potentials such as Quarton and Fluxonium, and their Hamiltonians are extracted and solved to validate the transition frequencies and calculate the coherence times of both the original qubits and their Trainmon-based versions.

## Introduction

Since the advent of circuit quantum electrodynamics, many superconducting circuits have been introduced, each offering unique properties [1]. A particularly active area of research involves the development of new circuit architectures that are protected against various noise channels. Notable contributions include the transmon qubit [2], which offers insensitivity to charge noise, and the fluxonium qubit [3], which provides resilience against relaxation. Although these qubits differ in many aspects, they share a common foundation: the typical superconducting qubit circuit consists of three fundamental components—a Josephson junction, a capacitor, and an inductor—connected in parallel. The generic Hamiltonian for such a circuit is given by [4].

*Equation 1*
$$H = 4E_c(n - n_g)^2 - E_J \cos(\phi - \phi_\text{ext}) + \frac{1}{2}E_L \phi^2$$

Here, $n$ and $\phi$ are the charge and phase operators, respectively, which satisfy the commutation relation $[n, \phi] = i$. The charging energy of the capacitor is given by $E_c = e^2/2C$, and $E_J$ is the Josephson energy of the Josephson junction, associated with the tunneling of Cooper pairs through the junction barrier. The inductive energy is $E_L = (\Phi_0/2\pi)^2 L$, where $L$ is the inductance. Here, $\Phi_0 = h/2e$ is the magnetic flux quantum, and $h$ is Planck's constant. The parameter $n_g$ denotes the biased gate charge of the qubit, and $\phi_\text{ext} = \Phi_\text{ext}/\Phi_0$ is the reduced external phase, related to the external magnetic flux $\Phi_\text{ext}$ threading the loop formed by the Josephson junction and the inductor.

Although the circuit appears simple, different values of $E_C$, $E_J$, and $E_L$ define distinct regimes in qubit design, collectively forming the so-called "periodic table" of superconducting qubits [4]. Among these, the transmon stands out as a particularly successful architecture. It exhibits high coherence times and strong resistance to charge noise by operating in the regime $E_J \gg E_C$ with $E_L = 0$. The transmon's simplicity and robustness against charge fluctuations have made it a widely adopted choice in experimental quantum computing research.

Fluxonium is another successful qubit known for its strong resilience against energy relaxation and its favorable $T_1$-protection characteristics [5]. It operates in the heavy-flux regime, characterized by $E_J \gg E_L \approx E_C$. This regime leads to the localization of wavefunctions, which enhances protection against relaxation processes. In contrast, the Blochnium qubit is designed to operate in the light-flux regime, where $E_C > E_J \gg E_L$. While this regime offers improved insensitivity to flux noise, it sacrifices the $T_1$-protection afforded by the heavy-flux configuration [6].

Designing a single-node superconducting circuit containing only one capacitor, one Josephson junction, and one inductor involves fundamental trade-offs [1]. Such a minimal circuit cannot simultaneously meet all the requirements for an ideal qubit, particularly when balancing desirable operational characteristics with protection against various noise channels. To address this limitation, numerous multi-node qubit circuits have been proposed to enhance noise resilience. Notable examples include the 0–π qubit [7], KITE qubit [8], bifluxon [9], and a qubit protected by two-Cooper-pair tunneling [10], all of which theoretically exhibit exponential insensitivity to a range of noise sources.

The unique properties of each qubit circuit design stem from its underlying potential well landscape. Efforts have been made to develop generalized frameworks that can accommodate a wide variety of qubit types. One such example is the generalized flux qubit [11], which serves as a flexible platform capable of realizing multiple qubit architectures—including the tunable transmon, persistent-current flux qubit, capacitively shunted flux qubit, and fluxonium. This framework proves useful for optimizing qubits within desired regimes, particularly in terms of anharmonicity and coherence. Another promising approach involves the use of parallel branches, each containing two Josephson junctions in series with differing transparencies [12]. While these frameworks are conceptually effective and show potential, they have yet to be fully developed or widely adopted in practical qubit design.

In this paper, we introduce a method to reverse-engineer a potential well named Trainmon. This frameworks is applied to well-known qubits such as fluxonium and quarton qubit to create the Trainmon potential of them. We also solve the Hamiltonian of the Trainmon qubit and compared these results with the original qubits. We show that even with a limited number of branches, it is possible to reconstruct the potential with a small relative error in both potential and eigenenergies. Our method differs from previous approaches in terms of circuit design and handling negative coefficients of Josephson energies. Finally, we calculate the coherence time of the Trainmon qubit derived from fluxonium and discuss how dephasing time could be affected in the case of the Trainmon.

## Method

Considering the circuit shown in Figure 1, it consists of an infinite number of parallel branches, each composed of $n$ identical Josephson junctions with Josephson energy $E_J^n$, connected in series with a common shunting capacitor of capacitance $C$. Assuming zero magnetic flux, the phase $\phi$ can be assigned to the central node. Under these conditions, the Hamiltonian of the circuit can be written as Equation 2.

Equation 2
$$\mathcal{H} = 4E_c \hat{n}^2 - E_J^1 \cos(\phi) - 2E_J^2 \cos\left(\frac{\phi}{2}\right) - 3E_J^3 \cos\left(\frac{\phi}{3}\right) + \cdots = 4E_c \hat{n}^2 - \sum_{n=1}^{\infty} n E_J^n \cos\left(\frac{\phi}{n}\right)$$

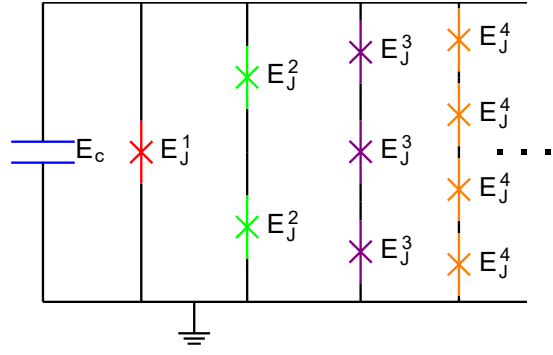

*Figure 1- schematic of Trainmon.*

It is important to note that Equation 2 is valid only under the condition $E_J^n/E_{c_{\text{eff}}} \gg 1$, where $E_{c_{\text{eff}}}$ is the total effective charging energy determined by the capacitance seen by the junctions [4]. Under this condition, phase slips can occur across the junctions, allowing the use of a quasi one-dimensional potential for each branch. If the junctions within each branch are identical, this approximation leads to the form of Equation 2.

The potential term of the Hamiltonian resembles a Fourier series, which can be used to reconstruct a given target potential. However, the basis functions $\cos(\phi/n)$ are not orthogonal over the range $[-\pi, \pi]$, which complicates direct analytical reconstruction. To address this, we employ optimization techniques—specifically, the `curve_fit` function from the `scipy.optimize` toolbox—to numerically determine the appropriate coefficients $E_J^n$ that best approximate the desired potential.

Some of the extracted Josephson energy coefficients are negative, which can be challenging to implement experimentally. To overcome this, one common approach is to shift all Josephson energy values by the absolute value of the most negative coefficient [12]. While this method is effective in eliminating negative energies, the global shift introduces a distortion in the potential that resembles a sinc function.

As an alternative, we apply an external magnetic flux to the corresponding loop to bias the qubit and shift the phase by $\pi$, thereby effectively generating negative Josephson energy coefficients.

In the presence of external flux, the fluxoid quantization condition can be applied to each superconducting loop [13], [14] as in Equation 3.

Equation 3 $\qquad \sum_{i=1}^{i \in l} \phi_i^l + \phi_e^l = 2\pi z, \quad z \in Z$

Here, $\phi_e^l$ is the reduced external phase threading loop $l$, and $\phi_i^l$ is the reduced phase drop across each Josephson junction in that loop.

Since each loop consists of two consecutive branches, and each branch in the loop $l$ contains $N_l \in \{n_1^l, n_2^l\}$ identical Josephson junctions, the phase drop across each junction is given by Equation 4 as depicted in Figure 2.

Equation 4 $\qquad \phi_i^l = \frac{\phi}{N_l} + \frac{\phi_{N_l}}{N_l}$

where $\phi_{N_l}/N_l$ is the total external flux effect in each branch, distributed equally among its junctions.

Substituting this into the fluxoid quantization condition of Equation 3 yields the Equation 5.

Equation 5 $\qquad \phi_{n_i} - \phi_{n_{i+1}} + \phi_e^l = 2\pi z, \quad z \in Z$

Under these conditions, the Trainmon Hamiltonian can be effectively reduced to a one-dimensional Hamiltonian, as shown in Equation 6 [11], [15].

Equation 6 $\quad H = 4E_c n^2 - \sum_{i=1}^{\infty} n_i E_J^{n_i} \cos\left(\frac{\phi}{n_i} + \frac{\phi_{n_i}}{n_i}\right)$

In the case of negative coefficients, an external magnetic flux should be applied such that it shifts the cosine potential of each junction by $\pi$, while still satisfying the fluxoid quantization condition.

Although an infinite number of branches would, in principle, provide a more accurate potential reconstruction, a limited number of branches can still yield a reasonable approximation over a specific range. Since this type of qubit is constructed by stacking branches of Josephson junctions, each contributing to the overall potential well—similar to the wagons of a train—we refer to it as Trainmon.

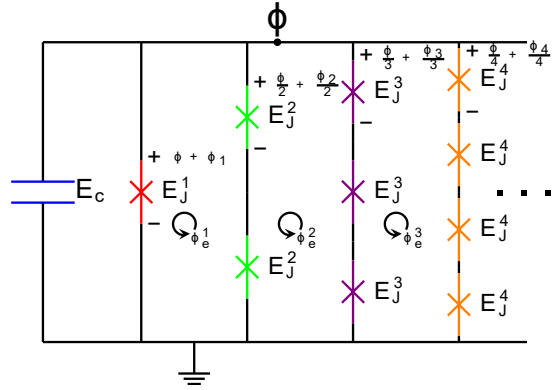

*Figure 2- Trainmon circuit considering the full external fluxes inside each loop.*

In this work, we test the Trainmon framework to realize a $\phi^4$ potential, as used in the Quarton qubit, which is known for exhibiting large anharmonicity.

## Results

We first tested the framework using the highly anharmonic Quarton qubit [11]. This qubit features a potential of the form of Equation 7.

Equation 7 $\quad U(\phi) = -\gamma E_J N \cos\left(\frac{\phi}{N}\right) - E_J \cos(\phi + \phi_e)$

where $\gamma$ is the ratio between the Josephson energy of the array junctions and that of the single distinct "black sheep" Josephson junction. Using the qubit parameters from Sample A [11], we reconstructed Trainmon versions of the Quarton potential over the range $[-\pi, \pi]$, employing configurations with one-, two-, and four-junction branches—referred to as the 124-Trainmon—each containing a different number of Josephson junctions. The resulting approximations are shown in Figure 3.

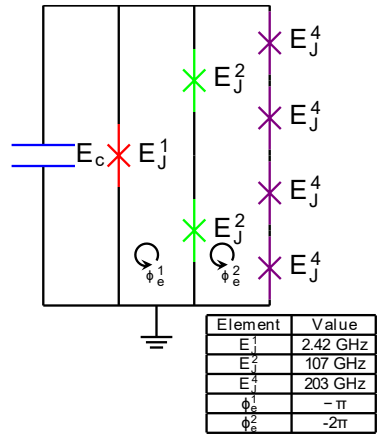

Figure 3- 124 Trainmon used to reconstruct the GFQ.

As shown in Figure 4-(a), the Quarton qubit potential closely overlaps with the Trainmon qubit potential depicted in Figure 4-(b), with the relative error between the two potentials reaching a maximum of only 0.02%. The potential range is another important factor to consider. In Figure 4-(c), the first three eigenenergies are plotted alongside the potential, indicating that the range $[-\pi, \pi]$ is appropriate, as it primarily encompasses the first two quantum states—those necessary for quantum computation with quartic potential characteristics.

The 124-Trainmon has an inherent periodicity of $8\pi$; therefore, the full potential is displayed over the range $[-4\pi, 4\pi]$. Within the range $[-\pi, \pi]$, however, the Trainmon most closely resembles the desired quartic potential, making it optimal for simulating the Quarton qubit behavior.

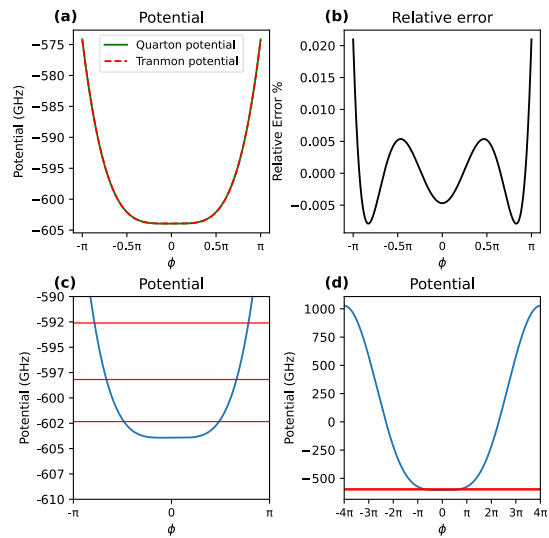

Figure 4-(a)Comparison of the Quarton potential and and the trainmon reconstructed potential of it. (b) For a better comparison the relative error of two potential is shown. (c) Three lowest eigenenergy of the Qurton potential(three solid red horizontal lines) is shown. (d) three lowest eigenenergies are shown in the whole range of the periodic potential of the trainmon in $[-4\pi, 4\pi]$.

In another attempt, a Trainmon version of the Fluxonium qubit was constructed. This Fluxonium circuit [16] is designed to operate in the heavy-flux regime, biased at $\Phi_{\text{ext}} = \Phi_0/2$, as illustrated in Figure 5-(a). Using the parameter values provided in the inset of Figure 5-(a), the Fluxonium potential was calculated over the range $[-3\pi, 3\pi]$ and used as a target for reconstruction with the 124-Trainmon configuration. The resulting potential, shown in Figure 5-(c) along with the corresponding Josephson energies in the inset, demonstrates excellent agreement.

The reconstructed Trainmon potential not only overlays nearly perfectly with the original Fluxonium potential within the target range but also maintains a correlation of 0.991 across the full periodic domain $[-4\pi, 4\pi]$, closely matching the behavior of the original circuit in Figure 5-(b).

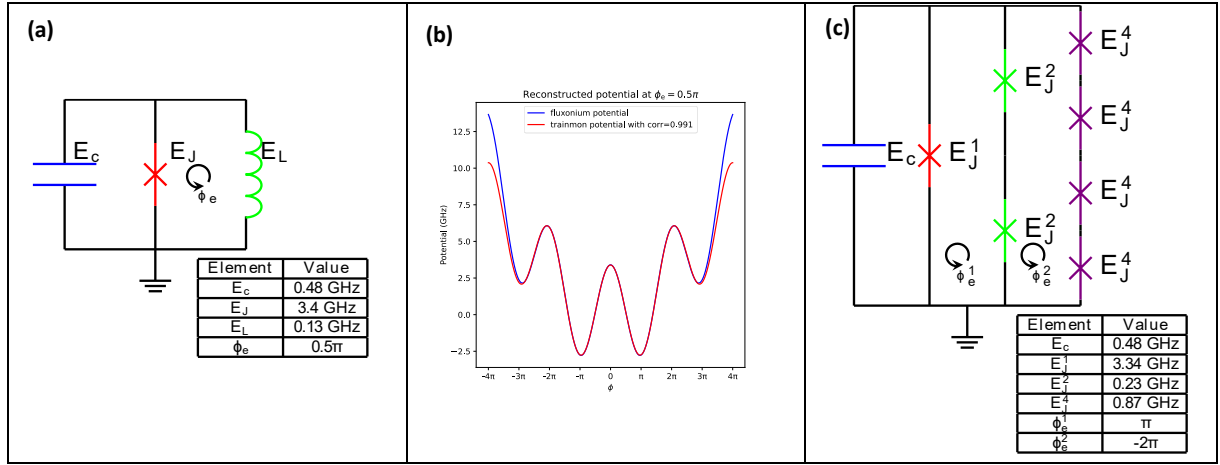

Figure 5-(a) schematic of the heavy fluxonium with the design values provided in inset. (b) comparison of the fluxonium potential and trainmon reconstructed potential in the $[-4\pi, 4\pi]$. (c) Trainmon schematic used to mimic the fluxonium potential with the design parametrs in insets.

As shown in Figure 5-(c), the Trainmon circuit includes two loops that allow external magnetic fluxes to penetrate and bias the circuit accordingly. For the Fluxonium configuration, the external fluxes applied to the first and second loops are $\Phi_e^1 = \Phi_0$ and $\Phi_e^2 = -2\Phi_0$, respectively.

Although Figure 5-(b) demonstrates that the reconstructed potentials of Fluxonium and Trainmon closely match, a key question remains regarding their transition energies. To address this, the Trainmon Hamiltonian was derived and numerically evaluated as detailed in Appendix A. Since the Trainmon circuit consists only of Josephson junctions and capacitors, its Hamiltonian can be efficiently expressed and solved in the charge basis.

For comparison, the Fluxonium Hamiltonian was solved using the `scQubits` toolbox [17], [18] and the results are summarized in Table 1. Both the first and second transition energies show nearly identical values between the two qubit types. The transition energy differences, defined as $\Delta_{E_{ij}} = E_{ij}^{\text{Trainmon}} - E_{ij}^{\text{Fluxonium}}$, and the relative transition energy errors, defined as $\delta_{E_{ij}} = \Delta_{E_{ij}}/E_{ij}^{\text{Fluxonium}}$, are both negligible. Specifically, $\Delta_{E_{01}}, \Delta_{E_{12}} \ll 1$ GHz and $\delta_{E_{01}}, \delta_{E_{12}} \ll 1$, indicating strong agreement between the Trainmon and Fluxonium in terms of energy spectrum.

Table 1- Numerical results for transition energies $E_{01}$ and $E_{12}$ of the Fluxonium and its Trainmon version

|  | Fluxonium | Trainmon |
|---|---|---|
| $E_{01}$(GHz) | 0.01379629 | 0.01379590 |
| $E_{12}$(GHz) | 2.95651201 | 2.94332248 |
| $\Delta_{E_{01}}$ (GHz) | -3.933460e-07 | |
| $\delta_{E_{01}}$ | -2.851181e-05 | |
| $\Delta_{E_{12}}$ (GHz) | -1.318953e-02 | |
| $\delta_{E_{12}}$ | -4.481171e-03 | |

Since the 124-Trainmon has two loops, it is possible to bias each loop independently with different external flux values and solve the Hamiltonian to calculate the corresponding transition energies for dispersion analysis. The energy dispersions $E_{01}$ and $E_{12}$ are shown as 3D plots in Figure 6-(a) and (b), with their corresponding contour plots presented in Figure 6-(c) and (d). These figures illustrate how the transition energies vary with changes in the external fluxes applied to each loop. Calculating such dispersions is a crucial step in determining the coherence times of a qubit [2], [7].

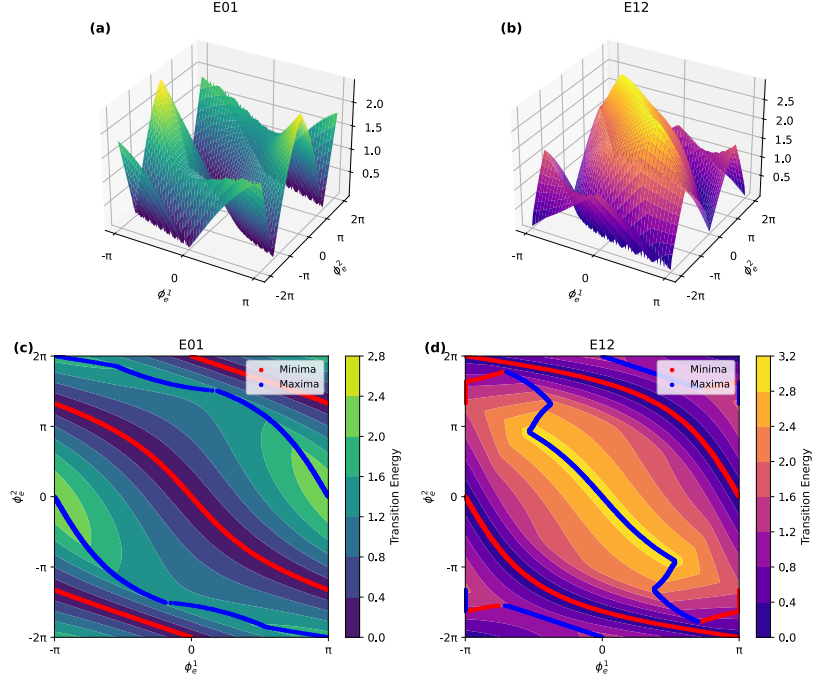

Figure 6-(a),(b) transition energy dispersion $E_{01}$ and $E_{12}$ as a function of exteral reduced fluxes $\phi_e^1$ and $\phi_e^2$. (c), (d) are the contour plot of the dispersion as a function of exteral reduced fluxes $\phi_e^1$ and $\phi_e^2$ with the local minimas and maximas as red and blue dots respectively. Dispersions calculated around the external operational biases of trainmon.

Since the 124-Trainmon, unlike Fluxonium, has two loops, it is more susceptible to flux noise. In this section, we calculate the dephasing time for both qubits to evaluate how Trainmon's coherence is affected by external flux fluctuations. We use Equation 8, which accounts for both the first and second derivatives of the energy dispersion with respect to the external fluxes in both loops [7], [19], [20], [21].

We assume a DC, spatially correlated $1/f$ flux noise model with a noise amplitude of $A_\Phi^{1/f} = 10^{-6}\Phi_0$, independently affecting each loop. For comparison, the dephasing time of the Fluxonium qubit is also calculated using the `scQubits` toolbox.

Equation 8
$$T_\varphi^\lambda = \left\{2A_\Phi^2 \left(\frac{\partial \omega_{ge}}{\partial \Phi}\right)^2 |\ln \omega_{ir} t| + 2A_\Phi^4 \left(\frac{\partial^2 \omega_{ge}}{\partial \Phi^2}\right)^2 \left[\ln^2\left(\frac{\omega_{uv}}{\omega_{ir}}\right) + 2\ln^2(\omega_{ir} t)\right]\right\}^{-1/2}$$

In Equation 8, $\omega_{ir}$ and $\omega_{uv}$ represent the infrared and ultraviolet frequency cutoffs, respectively, and $t$ is the experiment time. $A_\Phi$ denotes the noise amplitude. We set $\omega_{ir} = 2\pi \times 1$ Hz, $\omega_{uv} = 2\pi \times 3$ GHz, and $t = 10$ μs for our calculations.

The `scQubits` toolbox considers only the first derivative of the energy with respect to flux in calculating the dephasing time. To ensure a fair comparison, we computed both the first and second derivative contributions to the dephasing time for each loop of the Trainmon separately, as shown in Figure 7.

The calculated dephasing time for the first loop is $T_\phi^{\Phi_1} = 2821$ μs, and for the second loop $T_\phi^{\Phi_2} = 13739$ μs. The total dephasing time is then given by Equation 9.

Equation 9
$$\frac{1}{T_\phi^{\text{tot}}} = \frac{1}{T_\phi^{\Phi_1}} + \frac{1}{T_\phi^{\Phi_2}}$$

The Equation 9 yields $T_\phi^{\text{tot}} = 2340$ μs. For comparison, the dephasing time of the corresponding Fluxonium qubit, using the same parameters, is $T_\phi^{\text{Fluxonium}} = 2542$ μs. This indicates a modest reduction in dephasing time due to the additional loop structure in the Trainmon. However, the decrease is not significant enough to raise concern in designing a Trainmon version of the Fluxonium qubit.

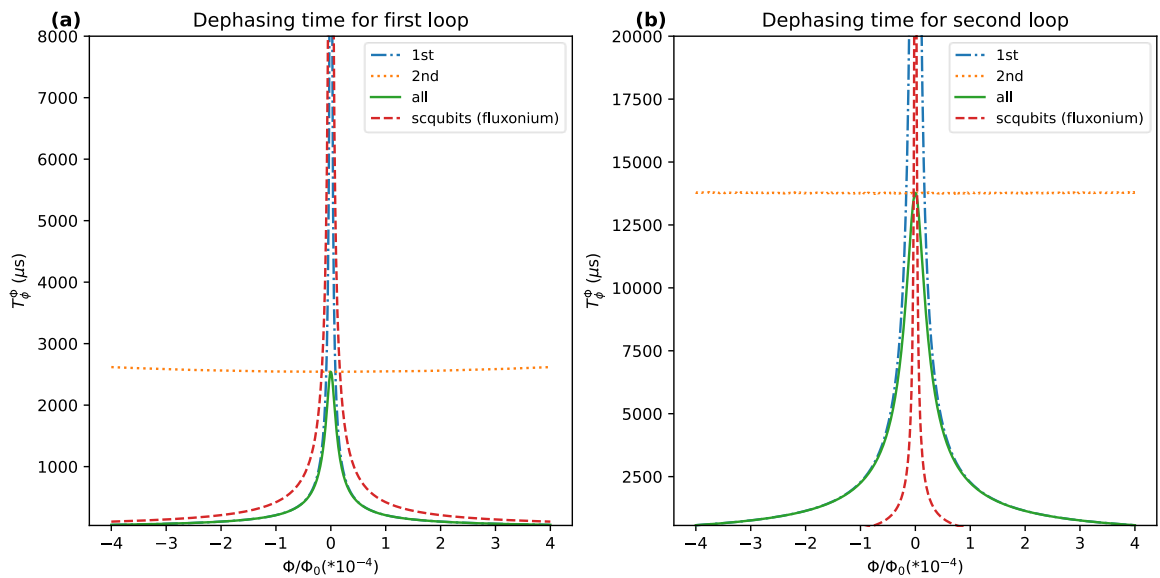

Figure 7-Dephasing time as a function of external reduced flux noise in the (a)first and (b)second loop. The blue dash-dot (-.) considering only the first derivative effect of dispersion while the orange dotted (:) line only considers the second derivative of the dispersion which is manly effective in the sweet-spot bias. The red dash (-) line is the `scQubits` result for the fluxonium. The green solid line is the overal dephasing time considering the first and second derivative of dispersion.

## Conclusion and outlook

A method is presented and tested across several qubit designs to implement a reverse engineering approach for superconducting qubits. This framework leverages parallel branches, each containing a different number of Josephson junctions, to contribute to the shaping of the qubit's quantum potential well. It offers flexibility in optimizing the potential landscape for specific purposes, such as enhancing noise insensitivity along selected noise channels.

# Appendix A

One of the main advantages of Trainmon is solving the Hamiltonian in charge space, the Traimon consist of a $\cos(\phi)$ element which is the josphson junction and a $\cos(\phi/n)$ elemtet representing an array of $n$ cascaded Identical Josephson junctions. The Hamiltonian for single Josephson junction potential in charge Hilbert space is $E_J \cos(\phi)|k\rangle = E_J/2(|k+1\rangle + |k-1\rangle)$ as for $\cos(\phi/n)$ element is $E_J \cos(\phi/n)|k\rangle = E_J/2(|k+1/n\rangle + |k-1/n\rangle)$, so for the potential of the form $U(\phi)|k\rangle = \sum_{n\alpha\{I\}} nE_J^n \cos(\phi/n + \phi_{ext}^n/n)$ that $I$ represents any set of $n$ Josephson junction arrays, the potential would have the following relationship in Equation 10.

Equation 10
$$U(\phi)|k\rangle = \sum_{n\alpha\{I\}} nE_J^n \left(\frac{1}{2}\cos\left(\frac{\phi_{ext}^n}{n}\right)\left[|k+\frac{1}{n}\rangle + |k-\frac{1}{n}\rangle\right] - \frac{1}{2j}\sin\left(\frac{\phi_{ext}^n}{n}\right)\left[|k+\frac{1}{n}\rangle - |k-\frac{1}{n}\rangle\right]\right)$$

By introducing a charge Hilbert space factors of $n = i \times \text{lcm}(I)$ which $i \propto \mathbb{N}$ and $\text{lcm}(I)$ is the least common multiple of set $\{I\}$, one could easily construct the Hamiltonian of the system in charge space and solve it to calculate the eigenenergies.

For example a 124-Trainmon depicted in Figure 3, has following set of branches $I = \{1,2,4\}$ which the least common multiple of set $I$ would be $\text{lcm}(I) = 4$. The choice of $i$ would be arbitrary depending on the resolution of the Hilbert space, in this example we consider $i = 1$, therefore we solve the potential in charges of the form $n \in \{\ldots, -2/4, -1/4, 0, 1/4, 2/4 \ldots\}$. By setting the limit for the charge space to $[-1,1]$ and considering zero external fluxes, the potential would take the form of Equation 11, This can be solved with QuTiP by setting the Josephson energies to unity and the external fluxes to zero for ease of observation. The main diagonal of the matrix represents the capacitive energy of each charge, which we ignore in this case. Each off-diagonal element represents the energy of the Josephson junction arrays.

Equation 11
$$U = \begin{pmatrix} 0.0 & -2.0 & -1.0 & 0.0 & -0.500 & 0.0 & 0.0 & 0.0 & 0.0 \\ -2.0 & 0.0 & -2.0 & -1.0 & 0.0 & -0.500 & 0.0 & 0.0 & 0.0 \\ -1.0 & -2.0 & 0.0 & -2.0 & -1.0 & 0.0 & -0.500 & 0.0 & 0.0 \\ 0.0 & -1.0 & -2.0 & 0.0 & -2.0 & -1.0 & 0.0 & -0.500 & 0.0 \\ -0.500 & 0.0 & -1.0 & -2.0 & 0.0 & -2.0 & -1.0 & 0.0 & -0.500 \\ 0.0 & -0.500 & 0.0 & -1.0 & -2.0 & 0.0 & -2.0 & -1.0 & 0.0 \\ 0.0 & 0.0 & -0.500 & 0.0 & -1.0 & -2.0 & 0.0 & -2.0 & -1.0 \\ 0.0 & 0.0 & 0.0 & -0.500 & 0.0 & -1.0 & -2.0 & 0.0 & -2.0 \\ 0.0 & 0.0 & 0.0 & 0.0 & -0.500 & 0.0 & -1.0 & -2.0 & 0.0 \end{pmatrix}$$